# Exact Solution of a Three-Dimensional Dimer System

H. Y. Huang,[1] V. Popkov,[2,3] and F. Y. Wu[1]

[1]*Department of Physics and Center for Interdisciplinary Research in Complex Systems,*
*Northeastern University, Boston, Massachusetts 02115*

[2]*Center for Theoretical Physics, Seoul National University, Seoul 151-742, Korea*

[3]*Institute for Low Temperature Physics, Kharkov, Ukraine (permanent address)*

## Abstract

We consider a three-dimensional lattice model consisting of layers of vertex models coupled with interlayer interactions. For a particular non-trivial interlayer interaction between charge-conserving vertex models and using a transfer matrix approach, we show that the eigenvalues and eigenvectors of the transfer matrix are related to those of the two-dimensional vertex model. The result is applied to analyze the phase transitions in a realistic three-dimensional dimer system.

05.50.+q

Typeset using REVTEX

An outstanding unsolved problem in the statistical mechanics of critical phenomena is the pursuit of exact solutions for realistic three-dimensional ($3D$) systems. While a large number of $2D$ systems have yielded to analyses [1], only a limited number of $3D$ systems have been solved. They include the $3D$ Ising model solved by Suzuki [2], the Zamolodchikov model [3] solved by Baxter [4] and its more recent $N$-state extension by Bazhanov and Baxter [5]. However, these models invariably suffer defects in one way or another: The Suzuki model turns out to be a $2D$ system in disguise, while the Zamolodchikov model and its extension involve unphysical negative Boltzmann weights. Similarly, a continuous string model in general $d$ dimensions solved by two of us [6] also involves negative weights. The solution of realistic physical $3D$ models has remained very much an open problem.

One approach toward solving realistic $3D$ models is to build from $2D$ systems. Indeed, such an approach has been suggested [7] and applied with some success recently [8,9]. However, in these considerations much attention has been placed to the algebraic structure of the transfer matrix and the associated Yang-Baxter equation, to the extent of masking the elegance of the solution. Here, we present a more general formulation, and deduce its solution via an alternate and yet much simpler consideration. The result is applied to analyze a realistic layered $3D$ dimer system.

Consider a simple-cubic lattice $\mathcal{L}$ of size $K \times M \times N$ with periodic boundary conditions. Regard $\mathcal{L}$ as consisting of $K$ copies of square lattices, of $M$ rows and $N$ columns each and stacked together as shown in Fig. 1. For simplicity, we shall speak of the square lattices as "layers" of $\mathcal{L}$. Label sites of $\mathcal{L}$ by indices $\{m, j, k\}$, with $1 \leq m \leq M$, $1 \leq j \leq N$ and $1 \leq k \leq K$. Within each layer of $\mathcal{L}$ define a $2D$ $q$-state vertex model whose lattice edges can be in $q$ distinct states. Label the state of the horizonal (vertical) edge incident at the site $\{m, j, k\}$ in the direction of, say, decreasing $\{m, j\}$ by $\alpha_{mjk}$ ($\beta_{mjk}$). It is convenient at times to suppress the subscripts $m$ and/or $k$ by adopting the notation

$$\beta_{m+1,j,k} \to \beta'_j, \qquad \beta_{m,j,k+1} \to \tilde{\beta}_j, \tag{1}$$

and similarly for the $\alpha$'s. Associate vertex weight $W_{mjk}$ to site $\{m, j, k\}$ which is a function of the configuration $C_{m,j,k} \equiv \{\alpha_{mjk}, \beta_{mjk}, \alpha_{m,j+1,k}, \beta_{m+1,j,k}\} \to \{\alpha_j, \beta_j, \alpha_{j+1}, \beta'_j\}$ of the four edges incident at the site $\{m, j, k\}$. Let the $\{m, j\}$ sites of two adjacent layers $k$ and $k+1$ interact with a Boltzmann factor $B_{mjk}$ which in the most general case is a function of the configurations $C_{m,j,k}$ and $C_{m,j,k+1}$ of the eight edges incident to the two sites. Then, the problem at hand is the evaluation of the partition function

$$Z_{MNK} = \sum_{\alpha_{mjk}} \sum_{\beta_{mjk}} \prod_{k=1}^{K} \prod_{m=1}^{M} \prod_{j=1}^{N} \left( B_{mjk} W_{mjk} \right) \quad (2)$$

where the summations are taken over all edge states $\alpha_{mjk}$ and $\beta_{mjk}$, and the per-site "free energy" for any $K$

$$f_K = K^{-1} \lim_{M,N \to \infty} (MN)^{-1} \ln Z_{MNK}. \quad (3)$$

*The transfer matrix*: The partition function (2) can be evaluated by applying a transfer matrix in the vertical direction. In a horizontal cross section of $\mathcal{L}$ there are $NK$ vertical edges. Let $\{\beta_m\} = \{\beta_{mjk} | 1 \leq j \leq N, 1 \leq k \leq K\}$, $1 \leq m \leq M$ denote the states of these $NK$ vertical edges, and define a $2^{NK} \times 2^{NK}$ matrix $\mathbf{T}$ with elements

$$T(\{\beta_m\}, \{\beta_{m+1}\}) = \sum_{\alpha_{mjk}} \prod_{k=1}^{K} \prod_{j=1}^{N} \left( B_{mjk} W_{mjk} \right), \quad 1 \leq m \leq M. \quad (4)$$

Then one has

$$Z_{MNK} = \sum_{\beta_{mjk}} \prod_{m=1}^{M} T(\{\beta_m\}, \{\beta_{m+1}\})$$
$$= \text{Tr } \mathbf{T}^M$$
$$\sim \Lambda_{\max}^M \quad (5)$$

where $\Lambda_{\max}$ is the largest eigenvalue of $\mathbf{T}$.

It is clear that we need to restrict considerations to models which are soluble when the interlayer interaction is absent, or $B_{mjk} = 1$. This leads us to build $3D$ systems from soluble $2D$ models. It is also clear that the interlayer interaction $B_{mjk}$ should be such that the overall interlayer factor $\prod_{m,j,k} B_{mjk}$ can be conveniently treated. For this purpose we restrict considerations to $2D$ charge-conserving models.

For definiteness let the labels $\alpha_{mjk}$ and $\beta_{mjk}$ take on a set $I$ of $q$ integral values. For example, one can take $I = \{+1, -1\}$ for $q = 2$ and $I = \{+1, 0, -1\}$ for $q = 3$. A 2D vertex model is charge-conserving if its vertex weights are non-vanishing only when

$$\alpha_j + \beta_j = \alpha_{j+1} + \beta'_j \quad \text{(charge conservation)} \tag{6}$$

holds at all sites. Examples of charge conserving models are the $q = 2$ ice-rule models [10], the $q$-state string model [11], the $q = 3$ Izergin-Korepin model [12], and others [13].

A direct consequence of the charge-conserving rule (6) is deduced by summing (6) from $j = 1$ to $j = N$. This yields

$$y_k = \frac{1}{N} \sum_{j=1}^{N} \beta_j = \frac{1}{N} \sum_{j=1}^{N} \beta'_j \tag{7}$$

showing that the quantity $-1 \leq y_k \leq 1$ is independent of $m$. (For ice-rule models this fact is well-known.)

Next one introduces the interlayer interaction [7–9]

$$B_{mjk} = \exp\left(h(\alpha_j \tilde{\beta}_j - \tilde{\alpha}_{j+1} \beta'_j)\right). \tag{8}$$

Since the negation of $h$ corresponds to a reversal of the layer numberings, without loss of generality we can take $h \geq 0$. We now show quite generally that the interlayer interaction (8) leads to a considerable simplification of the transfer matrix. Consider first the product

$$\prod_{j=1}^{N} B_{mjk} = \exp\left(h \sum_{j=1}^{N} (\alpha_j \tilde{\beta}_j - \tilde{\alpha}_{j+1} \beta'_j)\right). \tag{9}$$

Summing over (6), or $\alpha_i + \beta_i = \alpha_{i+1} + \beta'_i$, for $i = \{1, j-1\}$ and $i = \{j+1, N\}$ for the layer $k + 1$, one obtains, respectively, the identities

$$\alpha_j = \alpha_1 + \sum_{i=1}^{j-1} (\beta_i - \beta'_i), \quad j = 2, 3 \cdots, N$$

$$\tilde{\alpha}_{j+1} = \tilde{\alpha}_1 - \sum_{i=j+1}^{N} (\tilde{\beta}_i - \tilde{\beta}'_i), \quad j = 1, 2, \cdots, N-1 \tag{10}$$

where we have used $\tilde{\alpha}_{N+1} = \tilde{\alpha}_1$. Substituting (10) into (9) and making use of the identity $\sum_{j=2}^{N} \sum_{i=1}^{j-1} = \sum_{i=1}^{N-1} \sum_{j=i+1}^{N}$ in the first summation in (9), one arrives after a little algebra at

$$\prod_{j=1}^{N} B_{mjk} = \exp\Big(Nh(\alpha_1\tilde{y} - \tilde{\alpha}_1 y) + Nh[f(\beta,\tilde{\beta}) - f(\beta',\tilde{\beta}')]\Big), \tag{11}$$

where $f(\beta,\tilde{\beta}) \equiv \sum_{j=2}^{N} \sum_{i=1}^{j-1} \beta_i \tilde{\beta}_j$. The numerical factor $f(\beta,\tilde{\beta})$, which is defined for each fixed $m$, is cancelled in the further product

$$\prod_{m=1}^{M} \prod_{j=1}^{N} B_{mjk} = \prod_{m=1}^{M} \exp\Big(Nh(\alpha_{m,1,k} y_{k+1} - \alpha_{m,1,k+1} y_k)\Big). \tag{12}$$

As a result, only the conserved quantities $y_k$ and the state $\alpha_{m,1,k}$ of the extremites of a row of horizontal edges appear in the product (12). This leads us to rewrite the partition function (2) as

$$Z_{MNK} = \text{Tr}\,(\mathbf{T}^{\text{eff}})^M \tag{13}$$

where $\mathbf{T}^{\text{eff}}$ is a matrix with elements

$$T^{\text{eff}}(\{\beta_m\},\{\beta_{m+1}\}) = \sum_{\alpha_{mjk}} \prod_{k=1}^{K} \Big(e^{Nh\alpha_{m,1,k}(y_{k+1}-y_{k-1})} \prod_{j=1}^{N} W_{mjk}\Big). \tag{14}$$

The problem is now reduced to one of finding the largest eigenvalue of $\mathbf{T}^{\text{eff}}$. In fact, expression (11) shows that $\mathbf{T}^{\text{eff}}$ is related to $\mathbf{T}$ by a similarity transformation $\mathbf{T}^{\text{eff}} = \mathbf{S}\,\mathbf{T}\,\mathbf{S}^{-1}$ where $\mathbf{S}$ is diagonal. It follows that $\mathbf{T}$ and $\mathbf{T}^{\text{eff}}$ have the same eigenvalues, and their eigenvectors are related. The task is now considerably simpler since one needs only to keep track of the $2D$ system. The problem is solved if the eigenvalues of the transfer matrix for the $2D$ vertex model can be evaluated for fixed $y_k$ and $\alpha_{m,1,k}$.

*The ice-rule model*: To illustrate the usefulness of this formulation, we now apply it to layers of ice-rule model with vertex weights $\{\omega_1,\omega_2,...,\omega_6\}$ (for standard notations relevant to present discussions see, for example, [10]). Let $\alpha = +1$ $(-1)$ denote arrows pointing toward right (left), and $\beta = +1$ $(-1)$ arrows pointing up (down). Then one verifies that the charge-conserving condition (6) is satisfied with $y_k = 1 - 2n_k/N$, where $n_k$ is the number of down arrows in a row of vertical edges in the $k$th layer. Introducing next the interlayer interaction (8), the eigenvalues of the matrix (14) are obtained by applying a global Bethe ansatz consisting of the usual Bethe ansatz for each layer. The algebra is straightforward and one obtains

$$Z_{MNK} \sim \max_{1 \le n_k \le N} \prod_{k=1}^{K} [\Lambda_R(n_k) + \Lambda_L(n_k)]^M, \tag{15}$$

with

$$\Lambda_R(n_k) = e^{-2h(n_{k+1}-n_{k-1})} \omega_1^{N-n_k} \prod_{j=1}^{n_k} \left( \frac{\omega_3\omega_4 - \omega_5\omega_6 - \omega_1\omega_3 z_j^{(k)}}{\omega_4 - \omega_1 z_j^{(k)}} \right)$$

$$\Lambda_L(n_k) = e^{2h(n_{k+1}-n_{k-1})} \omega_4^{N-n_k} \prod_{j=1}^{n_k} \left( \frac{\omega_1\omega_2 - \omega_5\omega_6 - \omega_2\omega_4/z_j^{(k)}}{\omega_1 - \omega_4/z_j^{(k)}} \right), \tag{16}$$

where $\Lambda_R$ ($\Lambda_L$) refers to the eigenvalue for $\alpha_{m,1,k} = +1$ ($-1$) [10] and, for each $1 \le k \le K$, the $n_k$ complex numbers $z_j^{(k)}$, $j = 1, 2, \ldots, n_k$ are the solutions of the Bethe ansatz equations

$$e^{4h(n_{k+1}-n_{k-1})}(z_j^{(k)})^N = (-1)^{n_k+1} \prod_{i=1}^{n_k} \left( \frac{B(z_i, z_j)}{B(z_j, z_i)} \right), \quad j = 1, 2, \cdots, n_k \tag{17}$$

where $B(z, z') = \omega_2\omega_4 + \omega_1\omega_3 zz' - (\omega_1\omega_2 + \omega_3\omega_4 - \omega_5\omega_6)z'$. Note that the Bethe ansatz equation (17), which is obtained by imposing the cancellation of unwanted terms in the Bethe ansatz solution, differs from its usual form (see, for example, [14]) in the inclusion of the exponential factor involving $h$. Various special forms of this solution has been given previously [7–9].

*A dimer system with interlayer interactions*: We now consider a 3D lattice model consisting of layers of honeycomb dimer lattices. The dimers, which carry weights $u, v, w$ along the three honeycomb edge directions, are close packed within each layer and, in addition, interact between layers. For two dimers incident at the same $\{m, j\}$ site in adjacent layers, the interaction energy is given in Table I.

The 2D honeycomb dimer system can be formulated as a five-vertex model, namely, an ice-rule model with the weights [14,15]

$$\{\omega_1, \omega_2, \omega_3, \omega_4, \omega_5, \omega_6\} = \{0, w, v, u, \sqrt{uv}, \sqrt{uv}\}. \tag{18}$$

The 5-vertex model is defined on a square lattice of size $M \times N$ mapping to an honeycomb lattice of $2MN$ sites [14,15]. The mapping is such that the edge state $\alpha = +1$ ($\beta = +1$) corresponds to the presence of a $v$ ($u$) dimer. It can then be verified that the interlayer interaction given in Table I can be written precisely in the form of (8) [16], and therefore we can use the ice-rule model results.

Substituting (18) into (15), one obtains

$$Z_{MNK} \sim u^{MNK} \max_{1 \leq n_k \leq N} \prod_{k=1}^{K} \prod_{j=1}^{n_k} \left( \frac{w}{u} + \frac{v}{u} z_j^{(k)} \right)^M \tag{19}$$

with the Bethe ansatz solution

$$z_j^{(k)} = e^{i\theta_j} e^{2h(y_{k+1}-y_{k-1})}, \quad j = 1, 2, \cdots, n_k \tag{20}$$

where $e^{i\theta_j}$ are $n_k$ distinct $N$th roots of $(-1)^{n_k+1}$. For a given $n_k$, the factor inside the parentheses in (19) attends its maximum if the $\theta_j$'s lie on an arc crossing the positive real axis and extending from $-\pi(1-y_k)/2$ to $\pi(1-y_k)/2$. Using (3) this leads to the per-site free energy

$$f_K = \ln u + \max_{-1 \leq y_k \leq 1} \frac{1}{K} \sum_{k=1}^{K} \frac{1}{2\pi} \int_{-\pi(1-y_k)/2}^{\pi(1-y_k)/2} \ln\left( \frac{w}{u} + \frac{v}{u} e^{2h(y_{k+1}-y_{k-1})} e^{i\theta} \right) d\theta. \tag{21}$$

This is our main result.

We have carried out analytic as well as numerical analyses of the free energy (21) for $K = 3 \times$ integer. Here we summarize the findings. For $h = 0$, the layers are decoupled and the property of the system is the same as that of the 2D system [14,15]. For large $h$, it is readily seen from Table I that the energetically preferred state is one in which each layer is occupied by one kind of dimers, $u$, $v$, or $w$, and the layers are ordered in the sequence of $\{w, v, u, w, v, u, \cdots\}$. It is also clear that for large $u$, $v$, or $w$, the system is also frozen with complete ordering of $u$, $v$, or $w$ dimers. These orderings are referred to as the $H$, $U$, $V$, and $W$ phases, respectively. The system can also be in two other phases. A $Y$ phase in which all layers have the same value of $y_k = y$ determined straightforwardly by maximizing (21), or

$$w^2 + v^2 + 2wv \cos\left( \frac{\pi}{2}(1-y) \right) = u^2, \tag{22}$$

and an $I$ phase which is the $H$ phase with any of the $w$, $v$, or $u$ layers replaced by layers with $y_k = y$. If the $v$ layer is replaced by a $y$ layer so that the ordering is $\{w, y, u, w, y, u, \cdots\}$, for example, then $y$ is given by (22) with $v$ replaced by $ve^{4h}$.

The phase diagram is found to be symmetric in $w$, $v$, and $u$. It is then convenient to plot the phase diagram using the coordinates

$$X = \ln(v/w) \qquad Y = (\sqrt{3})^{-1} \ln(vw/u^2) \qquad (23)$$

so that any interchange of the three variables $w$, $v$, and $u$ corresponds to a $120^o$ rotation in the $\{X, Y\}$ plane. The phase diagram for $h < h_0 = 0.2422995...$ is the same as in Fig. 2a but without the $H$ regime. Increasing the value of $h$ one finds the $H$ phase appear in $h_0 < h < h_1 = 0.2552479...$ as shown in Fig. 2a. At $h = h_1$ the $I$ phase appears (Fig. 2b), with its region extending to infinity when $h$ reaches $h_2 = (\ln 3)/4 = 0.2746531...$ (Fig. 2c). When $h$ reaches $h_3 = 0.3816955...$ and higher, the $Y$ phase disappears completely as shown in Fig. 2d. All transitions are found to be of first-order except the transitions between the $\{U, V, W\}$ and $Y$ phases, and between the $I$ and $H$ phases, which are found to be of second-order with a square-root divergence in the specific heat.

In summary, we have presented the formulation of a general $3D$ lattice model, and applied it to solve a realistic dimer system. The analysis can be extended to include dimer-dimer interactions within each layer [14], and further-neighbor interlayer interactions. Details of the present and further analyses will be presented elsewhere. Work has been supported in part by NSF Grant DMR-9313648 and by INTAS Grants 93-1324 and 93-0633. One of us (VP) thanks Prof. D. Kim for discussions.

FIGURES

FIG. 1. A three-dimensional lattice model consisting of layered vertex models.

FIG. 2. Phase diagrams of the dimer system. (a) $h_0 < h < h_1$, (b) $h_1 < h < h_2$, (c) $h_2 < h < h_3$, (d) $h > h_3$.

TABLES

TABLE I. Interaction energy between two dimers incident at the same $\{m,j\}$ site of adjacent layers. For example, a $v$ dimer in the $k$th layer interacts with a $u$ dimer in the $(k+1)$th layer with an energy $\epsilon h$. Here $\epsilon = +1$ $(-1)$ if the site is in sublattice A (B).

| layer $k \to k+1$ | $w$ | $v$ | $u$ |
|---|---|---|---|
| $w$ | 0 | $-h$ | $h$ |
| $v$ | $h$ | 0 | $\epsilon h$ |
| $u$ | $-h$ | $\epsilon h$ | 0 |